\documentclass[conference,10pt,a4paper]{IEEEtran}
\usepackage[utf8]{inputenc}

\usepackage{stfloats}

\usepackage{amsmath}
\usepackage{amssymb}

\usepackage{booktabs}
\usepackage{multirow}

\usepackage{array}
\newcolumntype{P}[1]{>{\centering\arraybackslash}p{#1}}

\usepackage{siunitx}

\usepackage{fancyvrb}

\usepackage{cite}

\usepackage{pgfplots}
\definecolor{color0}{rgb}{0.12156862745098,0.466666666666667,0.705882352941177}
\definecolor{color1}{rgb}{1,0.498039215686275,0.0549019607843137}
\definecolor{color2}{rgb}{0.172549019607843,0.627450980392157,0.172549019607843}
\definecolor{color3}{rgb}{0.83921568627451,0.152941176470588,0.156862745098039}
\pgfplotsset{compat=1.16}

\usepackage{glossaries}
\setacronymstyle{long-short}
\newacronym{awgn}{AWGN}{additive white Gaussian noise}
\newacronym{cdf}{CDF}{cumulative distribution function}
\newacronym{ecdf}{ECDF}{empirical cumulative distribution function}
\newacronym{isi}{ISI}{intersymbol interference}
\newacronym{los}{LOS}{line-of-sight}
\newacronym{mimo}{MIMO}{multiple-input multiple-output}
\newacronym{miso}{MISO}{multiple-input single-output}
\newacronym{mui}{MUI}{multi-user interference}
\newacronym{nlos}{NLOS}{non line-of-sight}
\newacronym{pdp}{PDP}{power-delay-profile}
\newacronym{sinr}{SINR}{signal-to-interference-plus-noise ratio}
\newacronym{simo}{SIMO}{single-input-multiple-output}
\newacronym{sir}{SIR}{signal-to-interference ratio}
\newacronym{snr}{SNR}{signal-to-noise ratio}
\newacronym{tdd}{TDD}{time-division-duplex}

\usepackage{cite}
\usepackage{amsmath,amssymb,amsfonts}
\usepackage{algorithmic}
\usepackage{graphicx}
\usepackage{textcomp}
\usepackage{xcolor}
\def\BibTeX{{\rm B\kern-.05em{\sc i\kern-.025em b}\kern-.08em
    T\kern-.1667em\lower.7ex\hbox{E}\kern-.125emX}}

\begin{document}
\bstctlcite{IEEEexample:BSTcontrol}
\VerbatimFootnotes

\title{Fading Margins for Large-Scale Antenna Systems}
\author{
\IEEEauthorblockN{Jens Abraham\IEEEauthorrefmark{1}, Torbj\"orn Ekman\IEEEauthorrefmark{2}}
\IEEEauthorblockA{
    Department of Electronic Systems, Norwegian University of Science and Technology, Trondheim, Norway\\
    ORCID: \IEEEauthorrefmark{1}0000-0002-2640-0876, \IEEEauthorrefmark{2}0000-0002-3413-191X
}\thanks{testing}
}

\maketitle

\begin{abstract}
Mobile phone operators have begun the roll-out of 5G networks, deploying massive MIMO base stations.
Commercial product ranges start with 16 independent radio chains connected to a large-scale antenna system to exploit both
channel hardening and favourable propagation in order to obtain increased spectral efficiency.
In this work, the \glsdesc{cdf} describing the gain for large-scale antenna systems considering spatial and spectral diversity is evaluated empirically in terms of a fading margin and compared to an analytical maximum diversity reference system.
This allows for a simple investigation of the trade-off between deployment size and exploitation of channel hardening.
For the considered site-specific measurement data, little additional diversity is harvested with systems larger than 32 antenna elements.
\end{abstract}


\section{Introduction}

Massive \gls{mimo} has seen a lot of development since its conceptual advent in 2010 \cite{marzetta_noncooperative_2010}.
During the last decade, both theoretical and experimental work have  explored the limits of the approach.
Nowadays, operators can acquire commercially available base stations that implement some version of massive \gls{mimo}.

A more contemporary view of massive \gls{mimo} including a proper definition of a massive \gls{mimo} cellular network is provided in \cite{sanguinetti_toward_2020}.
Additionally, the authors highlight the fact that most literature has only considered spatially uncorrelated radio channels due to mathematical tractability.
Unfortunately, this approach neglects important aspects of the physical reality, which can lead to misleading conclusions.

Some attempts at building and using channel sounders and testbeds have been made to measure radio channels in some specific environments, e.g. \cite{chen_exploration_2016,harris_performance_2017}.
This work uses datasets of a large measurement campaign from 2016 \cite{shepard_understanding_2016} for practical demonstration.

In the following manuscript, we will present the connection between a link budget fading margin and channel hardening for an increasing number of antennas.
Only a single user is considered\footnote{A single user implies a reduction to a \gls{miso} / \gls{simo} system.} to investigate the best case without complication caused by \gls{mui}.
This explores an additional way of determining the scaling for large-scale antenna systems in addition to the work in \cite{gunnarsson_channel_2020, bjornson_massive_2017, abraham_power_2019}.

First, an \gls{ecdf} based fading margin is introduced, giving a measurement-based figure of merit in standard radio engineering terms.
Second, a maximum diversity reference channel is formed with an $N$-tap \gls{pdp} for independent base station antennas in a single user setting.
The final section presents an empirical procedure to evaluate the obtained channel hardening (spatial and spectral diversity) which both shows the actual scaling and highlights the difference to the uncorrelated reference channel.
Since the procedure shows the diminishing returns explicitly, it allows to assess the useful amount of antenna elements from a diversity perspective at a specific site.

\section{Fading Margin}

The fading margin describes the excess amount of power that a link budget has to provide to counteract fading events due to multipath propagation.
It is in the interest of the radio engineer to reduce the required excess power to optimise a radio link.
This reduction is beneficial due to energy savings and reduced system interference.

Large-scale antenna systems with phase steered beams, as used by massive \gls{mimo} systems, have multiple advantages compared to single antenna systems.
The directional gain is increased due to the array factor, whereas fading is less severe due to low probability that all antenna elements experience fading at the same time (channel hardening).
Moreover, inter-system interference (favourable propagation) as well as interference with other systems is reduced due to spatio-temporal focusing of power.

To study the channel hardening scaling behaviour, we use a fading margin $F_M(p)$ in logarithmic units for probability $p$, defined by
\begin{equation}
    F_M(p) = 10 \log_{10} \left(\frac{Q(0.5)}{Q(p)}\right) \label{eqn:fading_margin},
\end{equation}
where $Q(p)$ is the quantile function or inverse \gls{cdf} with corresponding \gls{cdf} $F(x)$ such that:
\begin{equation}
    Q(F(x)) = x.
\end{equation}

\begin{figure}[tbp]
    \centering
\begin{tikzpicture}
\begin{semilogyaxis}[
tick align=outside,
tick pos=left,
xlabel={$|\mathfrak{h}|^2$ [dB]},
xmajorgrids,
xmin=-20, xmax=12,
ylabel={CDF},
ymajorgrids,
ymin=1e-4, ymax=1,
width=.47\textwidth, height=.23\textwidth
]
\addplot [semithick, color0] table {data/cdf_m_1_n_2.csv};
\draw (axis cs:-20,0.5) -- (axis cs:-0.761,0.5) -- (axis cs:-0.761,1e-4) node [anchor=south west] {$Q(0.5)$};
\draw (axis cs:-20,0.005)  node [anchor=south west] {$p$}  -- (axis cs:-12.861,0.005) -- (axis cs:-12.861,1e-4) node [anchor=south east] {$Q(p)$};
\draw[<->,thick] (axis cs:-12.861,1e-3)-- (axis cs:-0.761,1e-3) node [thick, midway, anchor=north] {$F_M(p)$};
\end{semilogyaxis}
\end{tikzpicture}    
    \caption{The fading margin $F_M(p)$ is defined with help of the \gls{cdf} of the channel gain. This example visualises $F_M(p=\num{5e-3}) = \SI{12.1}{\deci\bel}$ covering the fading between the median effective channel gain at $Q(0.5)$ and the target outage channel gain at $Q(p=\num{5e-3})$.}
    \label{fig:fading_margin_def}
\end{figure}
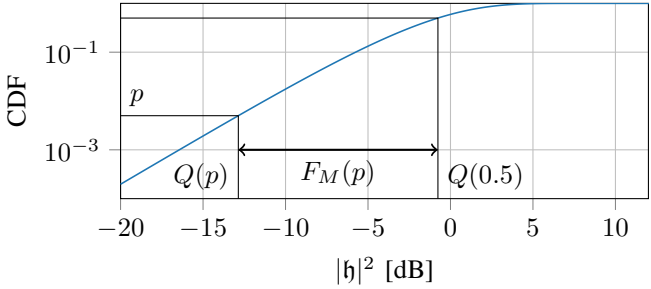

Furthermore, this fading margin is invariant to the array factor, which allows for comparison between different numbers of antenna elements.
The connection between this fading margin and the \gls{cdf} of the gain is visualised in Fig. \ref{fig:fading_margin_def}.
The steeper the \gls{cdf} the smaller becomes the fading margin.

Other definitions of a fading margin have been used in the literature.
The author of \cite{ramirez-mireles_performance_2001} motivates a fading margin as the difference between a fading channel and an \glsdesc{awgn} channel, whereas \cite{llano_analytical_2010} exchanges the median in \eqref{eqn:fading_margin} with the expected value of the channel gain as reference to calculate the fading margin.
By using the median gain as reference in \eqref{eqn:fading_margin}, a fading margin $F_M(p)$ of \SI{0}{\deci\bel} leads to half of the realisations falling short of and the other half exceeding it.
Furthermore, this fading margin can easily be extracted from \glspl{ecdf} and does not require any channel model.
In case of a symmetrical underlying fading distribution, both median and mean coincide.

\section{Channel Hardening}

After showing the fading margin, this section uses a tapped delay line model for the channel to explore channel hardening in both the spectral and spatial domain.
Time-reversal precoding \cite{oestges_characterization_2005} is applied to a maximum diversity reference channel.

Following an input-output description of a massive MIMO system in the downlink \cite{abraham_power_2019} and specialising it to the single user case gives:
\begin{equation}
    y[n] = \sqrt{\beta}~\mathfrak{h}[n] \star x[n] + e[n] \label{eqn:effective_channel_io}
\end{equation}
with symbols representing:
\begin{itemize}
    \item $\sqrt{\beta}$ - large-scale fading,
    \item $\mathfrak{h}[n]$ - effective downlink channel,
    \item $x[n]$ - transmitted signal,
    \item $y[n]$ - received signal,
    \item $e[n]$ - additive noise,
    \item $n$ - time index.
\end{itemize}
The large-scale fading coefficient can be estimated with the sample mean from raw channel measurements over a coherent block of base station antennas, channel taps and timestamps. 
The effective channel $\mathfrak{h}[n]$ is constituted by the sum of contributions from each antenna:
\begin{equation}
    \mathfrak{h}[n] = \sum_{m=1}^M h_{m}[n] \star w_{m}[n]  \label{eqn:eff_channel}
\end{equation}
convolving $h_{m}[n]$ and $w_{m}[n]$, being the uplink channel taps and precoding filter for antenna $m$, respectively.
In this paper we consider the commonly used time reversal weights for precoding, normalized to enforce unit gain of the effective channel: 
\begin{equation}
    w_{m}[n] = \frac{h_m^*[-n]}{\sqrt{\sum_{m=1}^M \sum_{n=1}^N |h_{m}[n]|^2}}, \label{eqn:weights}
\end{equation}
where $M$ and $N$ are the number of base station antennas and the number of taps of the tapped delay line model, respectively.

Under the assumption of uncorrelated effective channel taps, the instantaneous \gls{sinr} is:
\begin{equation}
    \gamma[n] = \frac{ |\mathfrak{h}[0]|^2}{ \sum_{l=-N,l\neq0}^{N} |\mathfrak{h}[l]|^2 + \frac{1}{\Gamma}} \label{eqn:sinr}
\end{equation}
where $\Gamma = \beta P_x / P_e$ is the mean \gls{snr} for transmit power $P_x$ and noise power $P_e$.
The interference term consists only of \gls{isi} (derivation see appendix \ref{app:sinr}).
For a multiuser \gls{mimo} discussion, the \gls{mui} would need to be added to the denominator.

The instantaneous \gls{sinr} is proportional with the numerator, showing the central role of the zero-delay tap $\mathfrak{h}[0]$ of the effective channel.
The \gls{isi} is captured in the off-centre taps of the effective channel in the denominator, as well as the noise influencing the mean \gls{snr}.
As expected, low \gls{snr} will lead to the noise limitation of the \gls{sinr}, whilst high \gls{snr} gives the interference limited regime.

It should be noted that the off-centre taps add up non-coherently, whilst $\mathfrak{h}[0]$ results from a coherent addition.
Hence, the offset between them is growing with an increasing number of independent base station antennas \cite{ghiaasi_effective_2019}.

Solving the convolution in \eqref{eqn:eff_channel} for the zero-delay results in
\begin{equation}
    \mathfrak{h}[0] = \sqrt{\sum_{m=1}^M \sum_{n=1}^N |h_{m}[n]|^2}. \label{eqn:zd_tap}
\end{equation}
Hence, both independent taps and antennas are increasing the instantaneous \gls{sinr}, where the number of taps is given by the propagation environment and bandwidth, whereas the number of antennas can be adjusted to improve the link.

To explore the scaling of the fading margin with respect to the number of antennas, an artificial reference channel can be considered.
The best case from a diversity perspective would be an $N$ independent tap channel with equal mean tap power ($1/N$).
This ensures unit gain and is in line with $\sqrt{\beta}$ capturing large-scale fading.
For a rich scattering environment, these channel taps can be modelled by complex normal random variables with Rayleigh distributed amplitudes.
Note, that $N$ represents the spectral diversity of the radio environment.

Both independent taps and independent antennas contribute in the same manner to $\mathfrak{h}[0]$.
The power gain $|\mathfrak{h}[0]|^2$ of the zero-delay effective channel tap is a sum over independent squared Rayleigh variables.
Squared Rayleigh distributed random variables are exponentially distributed and their sum is Gamma distributed \cite{mathai_storage_1982} with shape $MN$ (as each tap per antenna contributes) and scale $1/N$:
\begin{equation}
    |\mathfrak{h}[0]|^2 \sim \Gamma(MN, 1/N) \label{eqn:tap_dist}.
\end{equation}
It follows that the squared zero-delay tap of the effective channel has mean $M$ and variance $M/N$:
\begin{align}
    \mathcal{E} \left\{ |\mathfrak{h}[0]|^2 \right\} &= M \\
    \mathcal{V} \left\{ |\mathfrak{h}[0]|^2 \right\} &= \frac{M}{N}.
\end{align}

We can show that the Gamma distribution fulfils the condition for channel hardening by inserting it into \cite[Eqn. (2.17)]{bjornson_massive_2017}.
Evaluation of this squared coefficient of variation:
\begin{equation}
    \frac{\mathcal{V} \left\{ |\mathfrak{h}[0]|^2 \right\}}{\left(\mathcal{E} \left\{ |\mathfrak{h}[0]|^2 \right\} \right)^2} = \frac{1}{N M} \label{eqn:scov}
\end{equation}
shows convergence towards zero for a growing number of antennas or channel taps.
The authors of \cite{bjornson_massive_2017} state that a squared coefficient of variation order of \num{e-2} or smaller is enough to obtain hardening in an uncorrelated setting.
Hence, the effective channel can exhibit channel hardening with 4 taps and 32 antenna elements at the base station.
Unfortunately, \eqref{eqn:scov} is not offering an easily interpretable quantification of channel hardening.
This gap can be filled with the fading margin approach, as shown in the rest of the manuscript.

\begin{figure}[tbp]
\centering
\begin{tikzpicture}
\begin{semilogyaxis}[
legend entries={{$M=1$},{$M=4$},{$M=16$},{$N=1$},{$N=4$},{$N=16$}},
legend style={at={(0.5,1.05)},anchor=south},
legend columns=3, 
tick align=outside,
tick pos=left,
xlabel={$|\mathfrak{h}|^2$ [dB]},
xmajorgrids,
xmin=-15, xmax=15,
ylabel={CDF},
ymajorgrids,
ymin=1e-4, ymax=1,
width=.47\textwidth, height=.23\textwidth
]
\addlegendimage{no markers, color0}
\addlegendimage{no markers, color1}
\addlegendimage{no markers, color2}
\addlegendimage{no markers}
\addlegendimage{no markers, dashed}
\addlegendimage{no markers, dotted}
\addplot [thick, color0] table {data/cdf_m_1_n_1.csv};
\addplot [thick, color0, dashed] table {data/cdf_m_1_n_4.csv};
\addplot [thick, color0, dotted] table {data/cdf_m_1_n_16.csv};
\addplot [thick, color1] table {data/cdf_m_4_n_1.csv};
\addplot [thick, color1, dashed] table {data/cdf_m_4_n_4.csv};
\addplot [thick, color1, dotted] table {data/cdf_m_4_n_16.csv};
\addplot [thick, color2] table {data/cdf_m_16_n_1.csv};
\addplot [thick, color2, dashed] table {data/cdf_m_16_n_4.csv};
\addplot [thick, color2, dotted] table {data/cdf_m_16_n_16.csv};
\end{semilogyaxis}
\end{tikzpicture}
\caption{Analytical \glspl{cdf} for a few configurations of the $N$-tap reference channel and an uncorrelated $M$ element antenna array. The colour indicates the number of antennas and the line style the number of channel taps. Both spectral and spatial aspects contribute to the steepness of the curve (diversity), but only antenna elements improve the array gain. E.g. the $M=4$, $N=1$ configuration shows the same outage probability as the $M=1$, $N=4$ configuration offset by the array gain.}
\label{fig:reference_cdfs}
\end{figure}
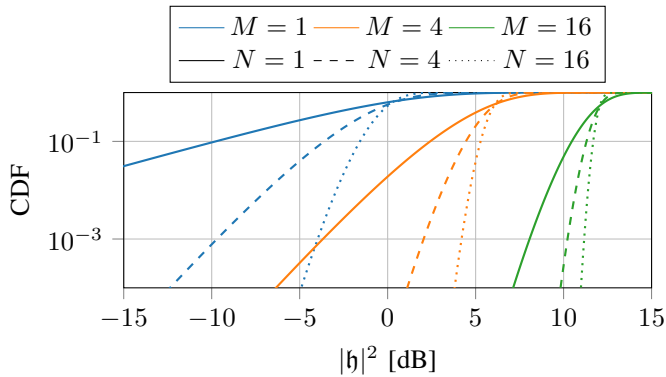

Fig. \ref{fig:reference_cdfs} shows a few \glspl{cdf} demonstrating the increasing steepness, giving a reduced fading margin, for growing number of antennas and taps.
The two lines corresponding to $MN = 4$ exhibit the same steepness and diversity.

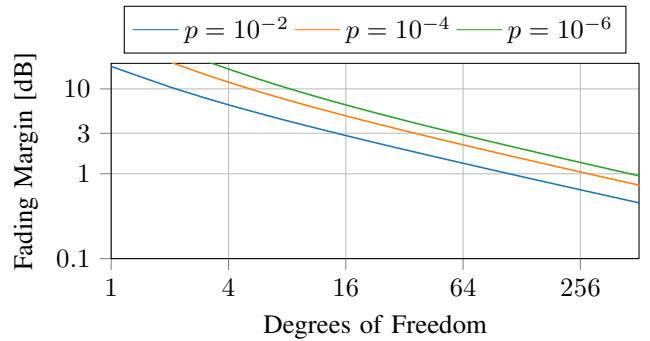
\begin{figure}[tbp]
\centering
\begin{tikzpicture}
\begin{axis}[
legend entries={{$p=10^{-2}$},{$p=10^{-4}$},{$p=10^{-6}$},},
legend style={at={(0.5,1.05)},anchor=south},
legend columns=3, 
tick align=outside,
tick pos=left,
xlabel={Degrees of Freedom},
xmode=log,
xtick={1,4,16,64,256},
xmajorgrids,
xmin=1, xmax=512,
ylabel={Fading Margin [dB]},
ymode=log,
ytick={.1,1,3, 10},
log ticks with fixed point,
ymajorgrids,
ymin=0.1, ymax=20,
width=.47\textwidth, height=.23\textwidth
]
\addlegendimage{no markers, color0}
\addlegendimage{no markers, color1}
\addlegendimage{no markers, color2}
\addlegendimage{no markers}
\addlegendimage{no markers, dashed}
\addlegendimage{no markers, dotted}
\addplot [semithick, color0] table {data/mf_p_0.01.csv};
\addplot [semithick, color1] table {data/mf_p_0.0001.csv};
\addplot [semithick, color2] table {data/mf_p_1.0e-6.csv};

\end{axis}
\end{tikzpicture}
\caption{Analytical fading margins for three different probabilities for the rectangular $N$-tap Rayleigh channel and $M$ independent antenna elements with $MN$ degrees of freedom. Qualitatively, the first few degrees of freedom improve the fading margin massively while additional ones have a reduced impact.}
\label{fig:fading_margin}
\end{figure}

Returning to the question how additional antennas in a large-scale antenna system can improve the fading margin, Fig. \ref{fig:fading_margin} shows the fading margin \eqref{eqn:fading_margin} versus degrees of freedom ($MN$) for different probabilities.
Taking a two tap channel for a single antenna system as a reference, gives a fading margin of about $\SI{30.7}{\deci\bel}$ to achieve an ultra-reliable outage probability of \num{e-6}.
Exploitation of channel hardening with 10 and 30 independent antennas would reduce the required fading margin ideally to \SI{5.6}{\deci\bel} and \SI{3.0}{\deci\bel}, respectively.
For later comparison to measurement data, fading margins for \num{e-3} are tabulated in Table \ref{tab:fm_gamma}.
It is obvious that the addition of more antennas has diminishing effects on the fading margin, whilst the array gain grows linearly.
The latter comes at the price of increased complexity for broadcast applications and user synchronisation, as we recently discussed in \cite{abraham_achievable_2020}.
The presented improvement of the fading margin is the best case result, since the model is based on independent antennas and uncorrelated channel taps for each antenna.
Real world systems would not see those improvements to the full extent.
Nonetheless, the qualitative behaviour helps to assess how many antennas are needed to improve the fading margin at a specific site.

\begin{table}[tbp]
    \centering
    \caption{Analytical fading margins at a probability of \num{e-3} for the rectangular independent $N$-tap Rayleigh channel and $M$ independent antenna elements.
    Increasing the degrees of freedom has diminishing returns.}
    \label{tab:fm_gamma}
    \begin{tabular}{lcccc}
        \toprule
        $F_M(10^{-3})$ & $M=1$ & $M=2$ & $M=4$ & $M=8$ \\
        \midrule
        $N=1$ & \SI{28.41}{\deci\bel} & \SI{15.68}{\deci\bel} & \SI{9.33}{\deci\bel} & \SI{5.9}{\deci\bel} \\
        $N=2$ & \SI{15.68}{\deci\bel} & \SI{9.33}{\deci\bel} & \SI{5.9}{\deci\bel} & \SI{3.88}{\deci\bel} \\
        $N=3$ & \SI{11.47}{\deci\bel} & \SI{7.09}{\deci\bel} & \SI{4.6}{\deci\bel} & \SI{3.08}{\deci\bel} \\
        $N=4$ & \SI{9.33}{\deci\bel} & \SI{5.9}{\deci\bel} & \SI{3.88}{\deci\bel} & \SI{2.62}{\deci\bel} \\
        \bottomrule
    \end{tabular}
\end{table}

To summarise, the behaviour of the fading margin caused by channel hardening for a changing number of base station antennas can be modelled for a tapped delay line massive \gls{mimo} channel.
Applying time-reversal precoding and relating the effective zero-delay channel coefficient to the instantaneous \gls{sinr} in the low average \gls{snr} regime under consideration of a rectangular reference channel, gives the best case for the evolution of the fading margin.
The actual fading margin in real world systems needs to be higher than the best case, due to spatial correlation, \gls{isi} and the reduced frequency diversity of non-rectangular non-Rayleigh channels.
Still, the system designer gets valuable insight into the scaling behaviour for base station antennas with respect to channel hardening.

\section{Case Study}

Considering that a large-scale antenna system at a specific site is supposed to be optimised, how to analyse the potential impact on the fading margin based on single antenna element measurements?
Ultimately, how many independent radio chains should the system support before the advantages are diminishing?
The general procedure for uncorrelated antennas is the following:
\begin{enumerate}
    \item Take single antenna multi carrier measurements over the array, spanning the largest deployable system on that site.
    \item Estimate a large-scale fading coefficient $\sqrt{\beta}$ for the base station and normalise the measurement data accordingly.
    \item Form an effective channel zero-delay tap $\mathfrak{h}[0]$ for each antenna position to determine the single element reference \gls{ecdf}.
    \item Form the effective channel for the array configurations in question.
    \item Evaluate the different fading margins for the sub-arrays to get an indication how large the optimised antenna array needs to be for a certain reliability requirement.
\end{enumerate}

Four datasets from \cite{shepard_understanding_2016} are used to showcase the outlined investigation in both \gls{los} and \gls{nlos} indoor environments, namely:
\begin{itemize}
    \item RICE A\footnote{\tiny Dataset: \verb|ArgosCSI-96x8-2016-11-04-04-18-58_2.4GHz_continuous_mob_LOS|} - \SI{2.4}{\giga\hertz} - \gls{los} environment,
    \item RICE B\footnote{\tiny Dataset: \verb|ArgosCSI-96x8-2016-11-04-05-57-41_2.4GHz_continuous_mob_NLOS|} - \SI{2.4}{\giga\hertz} - \gls{nlos} environment,
    \item RICE C\footnote{\tiny Dataset: \verb|ArgosCSI-96x8-2016-11-03-06-10-35_5GHz_continuous_mobile_LOS|} - \SI{5}{\giga\hertz} - \gls{los} environment,
    \item RICE D\footnote{\tiny Dataset: \verb|ArgosCSI-96x8-2016-11-03-04-36-53_5GHz_continuous_mob|} - \SI{5}{\giga\hertz} - \gls{nlos} environment.
\end{itemize}
For each dataset, a channel trace for user one is extracted, considering 14000 timestamps and 52 subcarriers over \SI{20}{\mega\hertz} bandwidth.
A maximum of 93 antenna elements is used, since antennas 17, 33 and 68 were providing much lower average signals in some datasets.

\begin{table*}[tbp]
    \centering
    \caption{Empirical fading margins for four different datasets \cite{shepard_understanding_2016} and different array configurations. 
    The improvement of the fading margin, due to usage of spectral diversity, is diminishing for a growing antenna array.}
    \label{tab:fm_real_world}
    \begin{tabular}{ccccccccc}
        \toprule
        & \multicolumn{8}{c}{Fading Margin $F_M(10^{-3})$} \\
        \cmidrule(lr){2-9}
        & \multicolumn{2}{c}{Single Element} & \multicolumn{2}{c}{Array (8 elements)} & \multicolumn{2}{c}{Array (32 elements)} & \multicolumn{2}{c}{Array (93 elements)}  \\
        \cmidrule(lr){2-3} \cmidrule(lr){4-5} \cmidrule(lr){6-7}  \cmidrule(lr){8-9}
        Dataset & narrowband & wideband & narrowband & wideband & narrowband & wideband & narrowband & wideband \\
        \midrule
        RICE A & \SI{28.75}{\deci\bel} &  \SI{9.97}{\deci\bel} & \SI{7.77}{\deci\bel} & \SI{5.21}{\deci\bel} & \SI{5.58}{\deci\bel} & \SI{4.67}{\deci\bel} & \SI{4.99}{\deci\bel} & \SI{4.4}{\deci\bel} \\
        RICE B & \SI{29.27}{\deci\bel} & \SI{13.39}{\deci\bel} & \SI{10.39}{\deci\bel} & \SI{7.91}{\deci\bel} & \SI{8.84}{\deci\bel} & \SI{7.66}{\deci\bel} & \SI{7.54}{\deci\bel} & \SI{5.84}{\deci\bel}\\
        RICE C & \SI{28.88}{\deci\bel} & \SI{10.46}{\deci\bel} & \SI{7.81}{\deci\bel} & \SI{5.17}{\deci\bel} & \SI{5.29}{\deci\bel} & \SI{4.05}{\deci\bel} & \SI{4.42}{\deci\bel} & \SI{3.61}{\deci\bel}\\
        RICE D & \SI{28.97}{\deci\bel} & \SI{12.71}{\deci\bel} & \SI{8.85}{\deci\bel} & \SI{7.07}{\deci\bel} & \SI{6.20}{\deci\bel} & \SI{5.25}{\deci\bel} & \SI{5.79}{\deci\bel} & \SI{4.84}{\deci\bel}\\
        
        \bottomrule
    \end{tabular}
\end{table*}

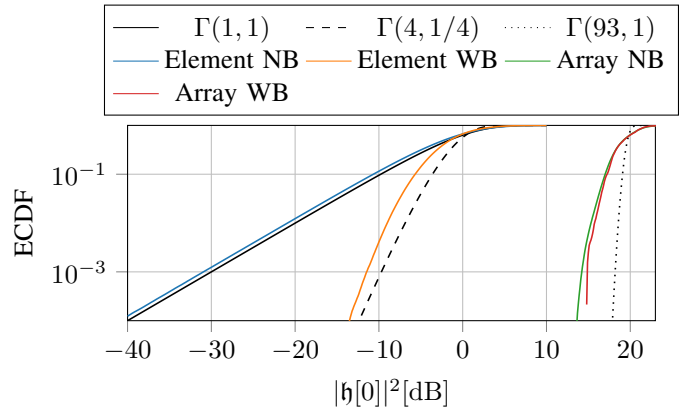
\begin{figure}[tbp]
\centering
\begin{tikzpicture}
\begin{semilogyaxis}[
legend entries={{$\Gamma(1,1)$},{$\Gamma(4,1/4)$},{$\Gamma(93,1)$},{Element NB},{Element WB},{Array NB},{Array WB}},
legend style={at={(0.5,1.05)},anchor=south},
legend columns=3, 
tick align=outside,
tick pos=left,
xlabel={$|\mathfrak{h}[0]|^2 [\si{\deci\bel}]$},
xmajorgrids,
xmin=-40, xmax=23,
ylabel={ECDF},
ymajorgrids,
ymin=1e-4, ymax=1,
width=.47\textwidth, height=.23\textwidth
]
\addlegendimage{no markers, black}
\addlegendimage{no markers, black, dashed}
\addlegendimage{no markers, black, dotted}
\addlegendimage{no markers, color0}
\addlegendimage{no markers, color1}
\addlegendimage{no markers, color2}
\addlegendimage{no markers, color3}

\addplot [semithick, black] table {data/gamma_1_1.csv};
\addplot [semithick, black, dashed] table {data/gamma_4_1_4.csv};
\addplot [semithick, black, dotted] table {data/gamma_93_1.csv};
\addplot [semithick, color0] table {data/ecdf_nb_s_los.csv};
\addplot [semithick, color1] table {data/ecdf_wb_s_los.csv};
\addplot [semithick, color2] table {data/ecdf_nb_f_los.csv};
\addplot [semithick, color3] table {data/ecdf_wb_f_los.csv};

\end{semilogyaxis}
\end{tikzpicture}
\caption{The \glspl{ecdf} for the 'RICE A' dataset (\SI{2.4}{\giga\hertz}, continuous mobility of user 1, 14000 timestamps) show the probabilities of the coherent channel gain for four different configurations. The  single antenna narrowband configuration follows the behaviour of a single tap Rayleigh channel very closely, whilst both full array configurations with 93 antennas fall even short of the single tap 93 antenna element model. This can be caused by spatial correlation reducing the harvested spatial diversity.}
\label{fig:rice_los}
\end{figure}

Fig. \ref{fig:rice_los} shows the \glspl{ecdf} for the RICE A dataset.
Considering single antenna elements on single subcarriers, shows that the \gls{ecdf} has the same slope and diversity as a single Rayleigh tap channel ($\Gamma(1,1)$).
The small offset in amplitude for the lower tail might arise from the assumption that all timestamps for each antenna and subcarrier belong to the same large-scale fading region.

Considering the wideband channel over single elements improves the fading margin and shows a steeper slope for the lower tail of the distribution.
Here, only spectral diversity is exploited and the offset between narrow- and wideband shows a large improvement of \SI{18.78}{\deci\bel} at a probability of $10^{-3}$.
The wideband channel behaves similar to a reference channel with 4 taps ($\Gamma(4,1/4)$) with a slightly larger offset.

Investigation of the full array for both cases shows that the spectral degrees of freedom play a reduced role as the spatial degrees of freedom kick in.
The full array wideband and narrowband cases exhibit almost equally steep \glspl{ecdf}.
Hence, the spectral diversity is consumed by the usage of spatial diversity.
The \glspl{ecdf} fall short of the fading margin behaviour for a 93 degrees of freedom reference channel and are closer to 12 degrees of freedom.
The loss in fading margin at probability $10^{-3}$ of \SI{2.9}{\deci\bel} could be due to correlated antennas. 

Table \ref{tab:fm_real_world} tabulates the fading margins for the named cases and intermediate array sizes.
For the RICE A dataset, there is almost no improvement between 32 element antenna arrays and the full 93 antenna array.
(NB, the larger arrays provide fewer realisations for the \glspl{ecdf} and should be interpreted carefully.)

None of the four datasets are achieving the theoretical fading margin of \SI{1.47}{\deci\bel} at $10^{-3}$ probability for 93 independent antenna elements and a single tap reference channel.
This indicates that the employed array is subject to non-diminishing spatial correlation.
The \SI{5}{\giga\hertz} traces show slightly better fading margins, most likely due to the increased antenna element distance of about one wavelength and lower antenna correlation.

The system improvements for a 93 element array over the 32 element arrays are mainly due to an increased array gain and less due to increased channel hardening.
The trade-off between base station complexity and performance should take this observation into account.
For the particularly highlighted measurements, distributing 32 antenna elements over the available array size appears to be a good compromise between the number of radio chains and the exploitation of channel hardening.
A potential extension to the current work is the analysis of permutations over the available antenna elements to give better performance than the usage of smaller and dense sub-arrays.

\begin{figure*}[!t]
\normalsize
\setcounter{equation}{11}
\begin{equation}
\mathcal{E} \left\{ \left| y[n] \right|^2\right\} = \mathcal{E} \left\{ \left|\sqrt{\beta}~\mathfrak{h}[n] \star x[n] + e[n] \right|^2 \right\} = \beta \left| \mathfrak{h}[0] \right|^2 P_x + \beta \sum_{l=-N, l \neq 0}^N \left| \mathfrak{h}[l] \right|^2 P_x + P_e \label{eqn:exp_rx_pwr}
\end{equation}
\begin{equation}
\gamma[n] = \frac{\beta \left| \mathfrak{h}[0] \right|^2 P_x}{\beta \sum_{l=-N, l \neq 0}^N \left| \mathfrak{h}[l] \right|^2 P_x + P_e} = \frac{\Gamma \left| \mathfrak{h}[0] \right|^2}{\Gamma \sum_{l=-N, l \neq 0}^N \left| \mathfrak{h}[l] \right|^2 + 1} = \frac{\left| \mathfrak{h}[0] \right|^2}{\sum_{l=-N, l \neq 0}^N \left| \mathfrak{h}[l] \right|^2 + \frac{1}{\Gamma}}. \label{eqn:sinr_derivation}
\end{equation}
\hrulefill
\end{figure*}

\section{Conclusion}

This paper has provided a definition of an alternative fading margin and used it to evaluate channel hardening in large-scale antenna systems.
A reference channel based on a rectangular $N$-tap Rayleigh \gls{pdp} demonstrates the ideal scaling behaviour for independent antenna elements under the most diverse channel conditions.
Measured channels of arrays can be easily used to evaluate site-specific diversity in both the spatial and spectral domain.
This gives system designers a tool to trade available diversity with system complexity by accounting for the number of independent radio chains.

The \SI{20}{\mega\hertz} indoor channel measurements at \SIlist{2.4;5}{\giga\hertz} show little difference in the fading margin between narrow- and wideband channels for large-scale antenna arrays.
An analysis of channel hardening in a system with 32 (correlated) antenna elements shows almost the same performance as that of a 93 element array. The fading margin shows diminishing returns with increasing number of antennas.

The investigation highlights that the diversity gains, measured using the fading margin at specific sites, can be evaluated with a relatively simple protocol.

\appendix

\subsection{Instantaneous Effective Channel \gls{sinr}}\label{app:sinr}

The instantaneous \gls{sinr} for the effective channel $\gamma[n]$ with respect to the average \gls{snr} $\Gamma = \beta \frac{P_x}{P_e}$ for powers $P_x = \mathcal{V}\left\{ x[n] \right\}$ and $P_e = \mathcal{V}\left\{ e[n] \right\}$ as variance of the transmit signal and noise signal, respectively, can be derived from \eqref{eqn:effective_channel_io} by taking the expectation over both signals as in \eqref{eqn:exp_rx_pwr}.
Here, intended signal, \gls{isi} and noise have been isolated for uncorrelated effective channel taps allowing to define the \gls{sinr} as in \eqref{eqn:sinr_derivation}.

\vfill

\bibliographystyle{IEEEtranDOI}
\bibliography{IEEEabrv,2021_icc.bib}

\end{document}